\documentclass[twocolumn,twoside,slac_two]{revtex4}
\usepackage{graphicx}
\usepackage{fancyhdr}
\def\units#1{\hbox{$\,{\rm #1}$}}
\def\degrees{\hbox{${}^\circ$}}

\begin{document}

\title{Fermi-LAT spectral analysis of Fermi, Planck, Swift and radio selected samples of AGN}

\author{C.~Monte\footnote{corresponding author, E-mail: claudia.monte@ba.infn.it}, S.~Rain\'o, F.~Gargano} 
\affiliation{Istituto Nazionale di Fisica Nucleare - Sezione di Bari, I-70126, Bari, Italy}
\author{S.~Cutini, D.~Gasparrini}
\affiliation{Agenzia Spaziale Italiana (ASI) Science Data Center, I-00044 Frascati (Roma), Italy}
\author{for the Fermi-LAT Collaboration}
\affiliation{\\}
\author{J.~Leon Tavares}
\affiliation{Aalto University Mets\"{a}hovi Radio Observatory, Mets\"{a}ovintie 114, FIN-02540 Kylm, Finland}
\author{ G.~Polenta}
\affiliation{Agenzia Spaziale Italiana (ASI) Science Data Center, I-00044 Frascati (Roma), Italy, \\
INAF - Osservatorio Astronomico di Roma, I-00044, Monte Porzio Catone (Roma), Italy}
\author{for the Planck Collaboration} 

\begin{abstract}
Blazars are jet-dominated extragalactic objects characterized by the emission of strongly variable non-thermal radiation across the entire electromagnetic spectrum.  Therefore, the study of blazars (and in general of radio loud AGN) through the use of multi-frequency simultaneous data is essential in order to understand the physical processes that take place in these objects. With Planck, Fermi and Swift simultaneously on orbit, complemented with other space and ground-based observatories, it is possible to assemble high-quality multi-frequency simultaneous broad-band spectra of large and statistically well-defined samples of radio-loud AGN.
In particular, four samples of sources have been selected. The first three samples are flux limited in the high energy part of the electromagnetic spectrum: the soft X-ray ($0.1\div2 \units{keV}$) sample includes 43 sources from the Rosat All Sky Survey Bright Source Catalog, the hard X-ray ($15\div150\units{keV}$) sample includes 34 sources from the Swift-BAT 54 months source catalog and the gamma-ray sample includes 50 sources from the Fermi-LAT 3 months Bright AGN Source List. The fourth sample is radio flux limited, including 104 bright northern and equatorial radio-loud AGN (most of which have been monitored at Metsahovi Radio Observatory for many years) with average radio flux density at $37\units{GHz}$ greater than $1\units{Jy}$.
We present the methods applied and the results of the analysis performed using Fermi-LAT data for all sources in the four different samples of AGN.
\end{abstract}

\maketitle

\thispagestyle{fancy}

\section{Introduction}
Blazars are well-known jet-dominated extragalactic objects characterized by the emission of strongly variable and polarized non-thermal radiation across the entire electromagnetic spectrum, from radio to high energy $\gamma$-rays. The strong emission of blazars at all wavelengths makes them the dominant type of extragalactic sources in the radio, $\mu$-wave, and $\gamma$-ray bands where the accretion and other thermal emission processes do not produce significant amounts of radiation.

For these reasons blazars are hard to find at optical and X-ray frequencies, while dominating the $\mu$-wave and $\gamma$-ray high Galactic latitude sky. 
The advent of the Fermi \cite{key:atwood} and, more recently, of the Planck satellite \cite{key:tauber},  capable of probing deeply these two last observing windows, combined with the versatility of the Swift observatory \cite{key:gehrels}, and the observations by a number of ground based observatories, is giving us the unprecedented opportunity to collect multi-frequency data for very large samples of blazars in order to assemble simultaneous or quasi-simultaneous broad-band spectra.

\section{The samples}
In order to explore the blazars' parameters space from different viewpoints we have adopted different criteria to select the list of blazars to be observed simultaneously by Planck, Swift and Fermi. 

The first three samples of blazars are flux limited in the high energy part of the electromagnetic spectrum \cite{key:planck1}.
{\em The soft X-ray flux limited sample}, including 43 sources, was defined starting from the Rosat All-Sky Survey Bright Source Catalog (1RXS \cite{key:voges}), selecting all the blazars with count rate larger than $0.3\units{counts/s}$ in the $0.1\div2.4\units{keV}$ energy band, and radio flux density $S_{5GHz}>200\units{mJy}$. {\em The hard X-ray flux limited sample}, including 34 sources, was defined starting from the $Swift$-BAT 54~month source catalog \cite{key:cusumano}, selecting all blazars with X-ray flux $>10^{-11}\units{erg~cm^{-2}~s^{-1}}$ in the  $15\div150\units{keV}$ energy band and radio flux density $S_{5GHz}>100\units{mJy}$.
{\em The $\gamma$-ray flux limited sample}, including 50 sources, was created starting from the $Fermi$-LAT Bright Source List \cite{key:abdo}, selecting all the high galactic ($\mid b\mid >$10$\degrees$) blazars detected with high significance ($TS>100$), with a flux cut $F(E>100\units{MeV})>8\times10^{-8}ph~cm^{-2}s^{-1}$ and radio flux density $S_{5GHz}>1\units{Jy}$.

The last sample \cite{key:planck2}, including 104 sources, is {\em the radio flux density limited sample} \cite{key:voges}: it includes all northern and equatorial radio-loud AGN with declination $\geq-10\degrees$ that have a measured average radio flux density $S_{37GHz}>1\units{Jy}$. 

\section{Fermi-LAT data analysis}
$Fermi$-LAT data taken from August~4,~2008 to November~4,~2010 have been analyzed selecting for each source in the previously defined samples \cite{key:planck1} \cite{key:planck2}, only photons above $100\units{MeV}$, belonging to the diffuse class \cite{key:atwood} which have a low probability of background contamination, within a $15\degrees$ Region of Interest (RoI) centered around the source itself. The data were analyzed with a binned maximum likelihood technique \cite{key:mattox} using the analysis software ({\em gtlike}) developed by the LAT team. A model accounting for the diffuse emission as well as for the nearby $\gamma$-ray sources is included in the fit. 

For the evaluation of the $\gamma$-ray SEDs, the whole energy range from $100\units{MeV}$ to $300\units{GeV}$ has been divided into 2 equal logarithmically spaced bins per decade. In each energy bin the standard gtlike binned analysis has been applied assuming a power law spectrum for all the point sources in the model, with photon index fixed to 2. The flux of the source in all selected energy bins has been evaluated, requiring in each bin a Test Statistics (TS) greater than 10 and the ratio between the flux and its error greater than 0.5. If these conditions were not satisfied, an upper limit (UL) has been evaluated in that energy bin. 
For each source in the four samples, three different integration periods have been considered: 
\begin{itemize}
\item{simultaneous observations (data accumulated during the period of Planck observation of the source);}
\item{quasi-simultaneous observations (data integrated over a period of two months centered on the Planck observing period of the source);}
\item{27 month Fermi integration (data integrated over a period of 27 months from August 4, 2008 to November 4, 2010).}
\end{itemize}

\section{Spectral Energy Distributions}
\begin{figure*}
\centering
\includegraphics[width=130mm]{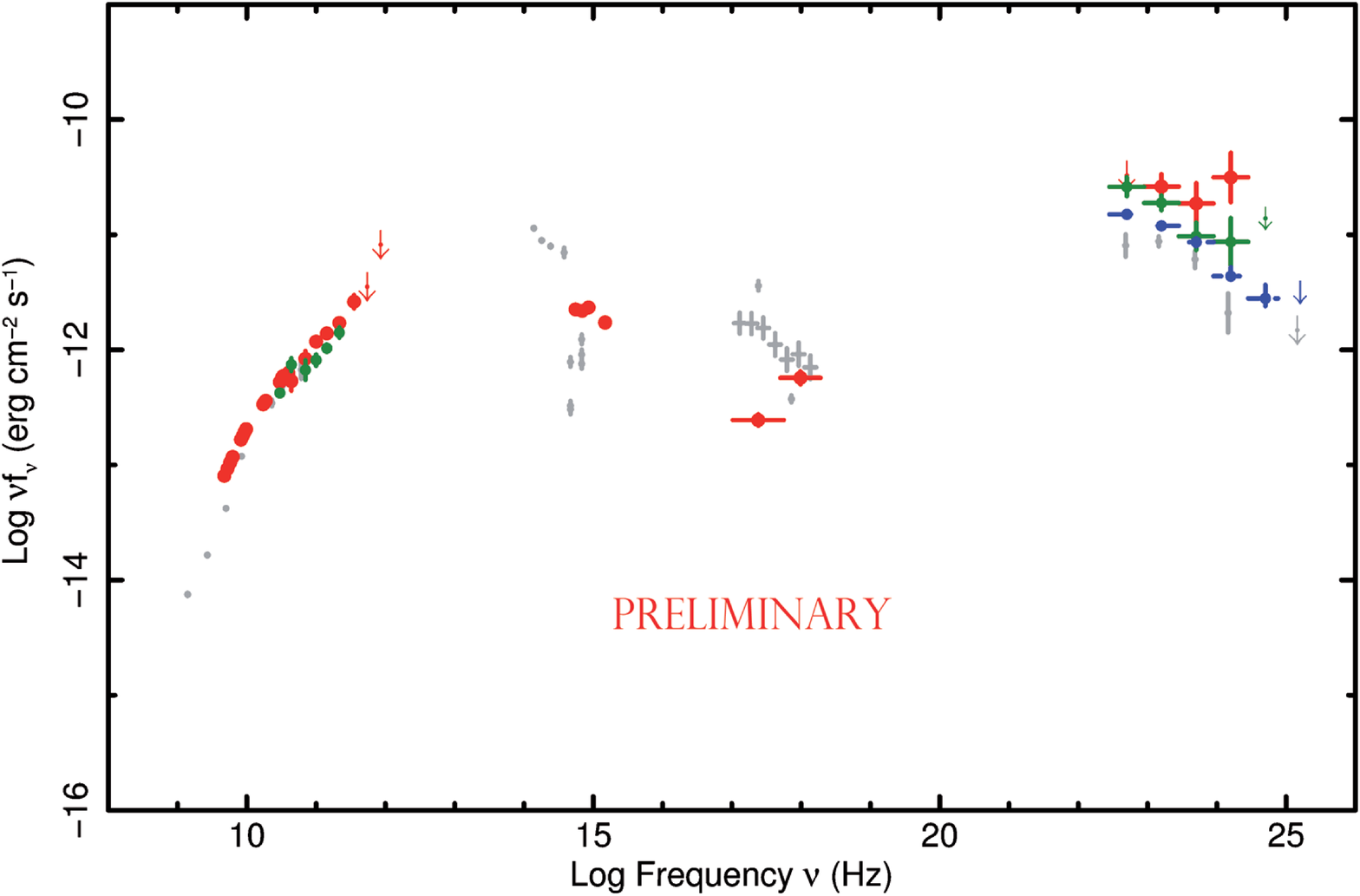}
\caption{SED of PKS 1124-186 from the soft X-ray sample \cite{key:planck1}. Red filled points: simultaneous multi-frequency data; green points: $\gamma$-ray data integrated over a period of 2 months centered on the Planck observing period, or ground-based data taken quasi-simultaneously; blue points: $Fermi$-LAT data integrated over 27 months; light gray points: literature or archival data.}
\label{fig:sed1}
\end{figure*}

\begin{figure*}
\centering
\includegraphics[width=130mm]{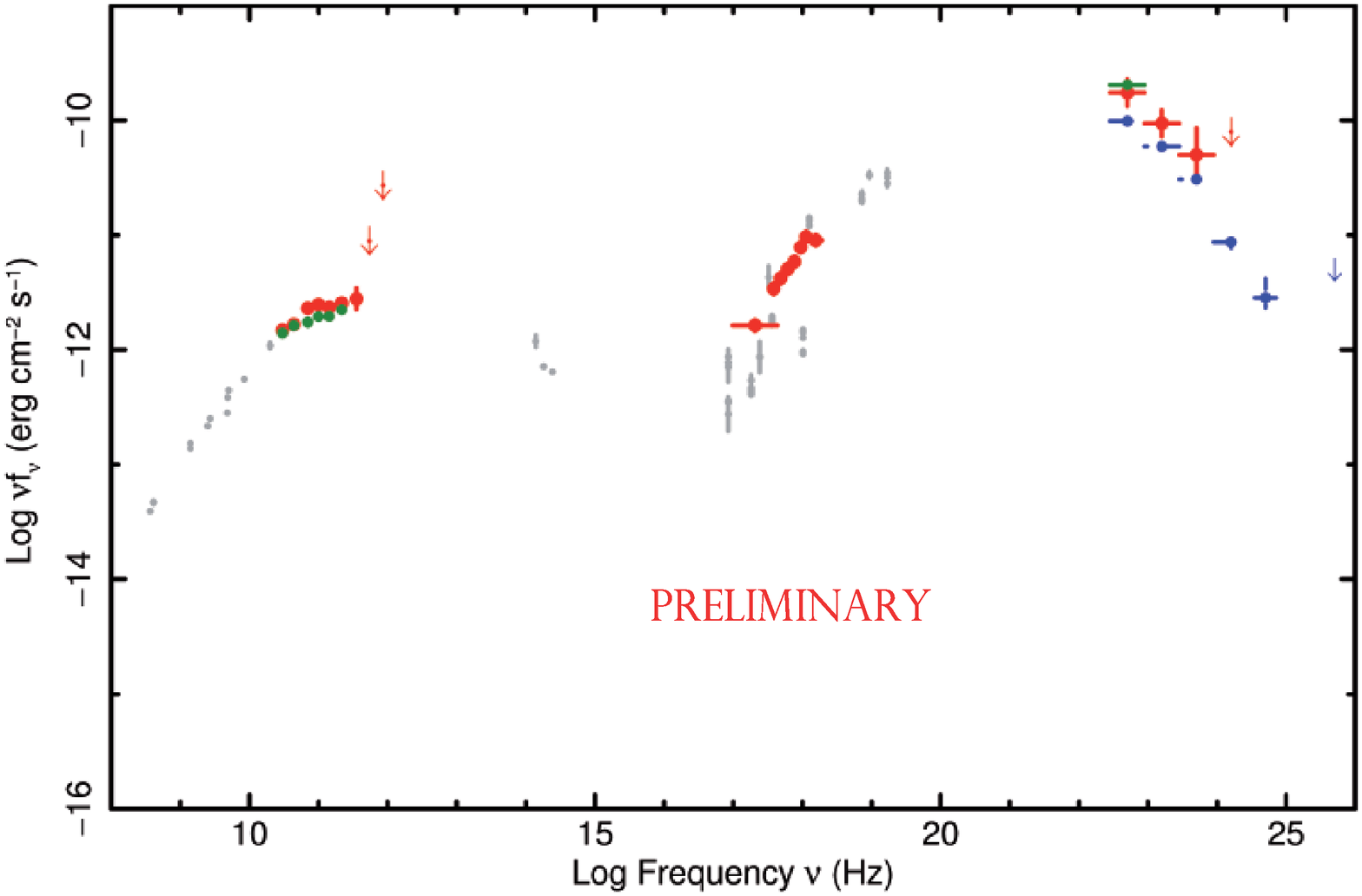}
\caption{SED of PKS B1830-210 from the hard X-ray sample \cite{key:planck1}. Red filled points: simultaneous multi-frequency data; green points: $\gamma$-ray data integrated over a period of 2 months centered on the Planck observing period, or ground-based data taken quasi-simultaneously; blue points: $Fermi$-LAT data integrated over 27 months; light gray points: literature or archival data.}
\label{fig:sed2}
\end{figure*}

\begin{figure*}
\centering
\includegraphics[width=130mm]{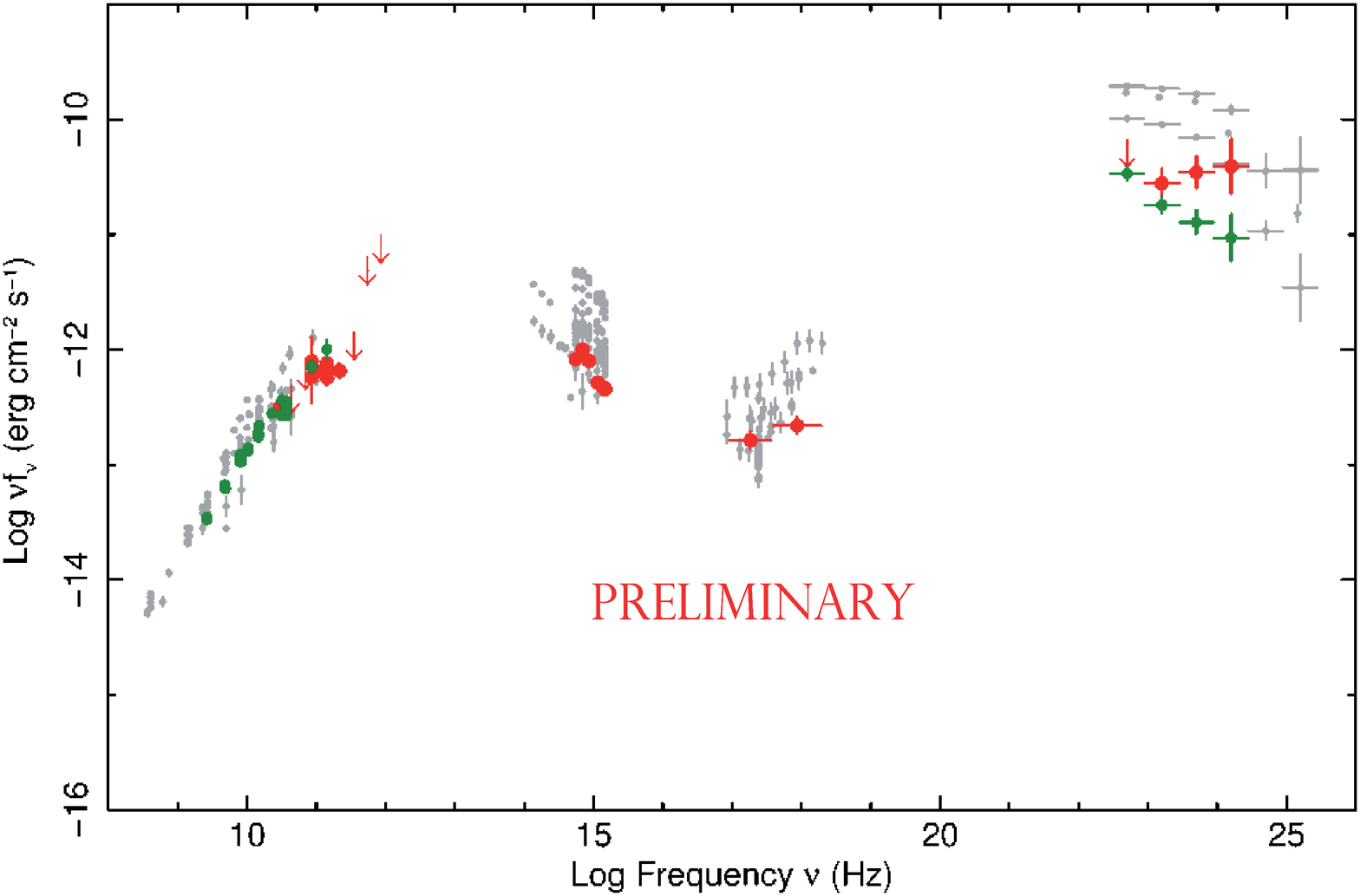}
\caption{SED of PKS 1502+106 from the $\gamma$-ray sample \cite{key:planck1}. Red filled points: simultaneous multi-frequency data; green points: $\gamma$-ray data integrated over a period of 2 months centered on the Planck observing period, or ground-based data taken quasi-simultaneously; blue points: $Fermi$-LAT data integrated over 27 months; light gray points: literature or archival data.}
\label{fig:sed3}
\end{figure*}

\begin{figure*}
\centering
\includegraphics[width=130mm]{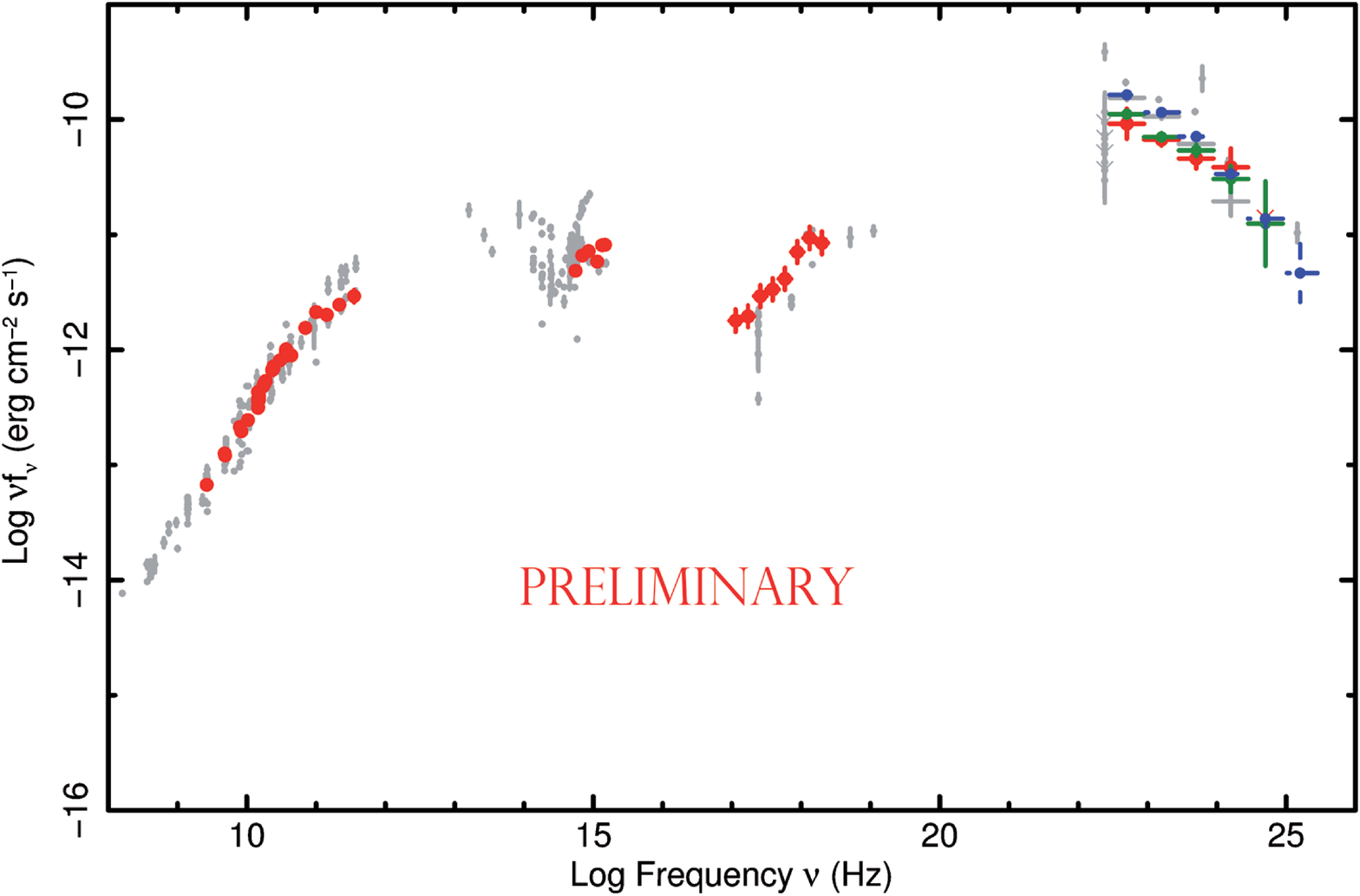}
\caption{SED of PKS 1510-089 from the radio sample \cite{key:planck2}. Red filled points: simultaneous multi-frequency data; green points: $\gamma$-ray data integrated over a period of 2 months centered on the Planck observing period, or ground-based data taken quasi-simultaneously; blue points: $Fermi$-LAT data integrated over 27 months; light gray points: literature or archival data.}
\label{fig:sed4}
\end{figure*}

The plot of radio to $\gamma$-ray flux distributions in the $Log\nu - Log\nu F_{\nu}$, widely known as a Spectral Energy Distribution (SED), is a powerful method of studying the physics of blazars.
The SEDs of PKS~1124-186 (from the soft X-ray sample), PKS~B1830-210 (from the hard X-ray sample), PKS~1502+106 (from the $\gamma$-ray sample) and PKS~1510-089 (from the radio sample) have been reported respectively in Fig.~\ref{fig:sed1}, Fig.~\ref{fig:sed2}, Fig.~\ref{fig:sed3}, Fig.~\ref{fig:sed4}. 
In all plots, red filled points (or UL) show simultaneous multi-frequency data, green points (or UL) show $\gamma$-ray data integrated over a period of 2 months centered on the Planck observing period, or ground-based data taken quasi-simultaneously and blue points (or UL) show $Fermi$-LAT data integrated over 27 months; literature or archival data are shown in light gray.

\section{Conclusions}
We have collected Planck, Swift, Fermi and ground based simultaneous multi-frequency data for a great number of blazars included in four statistically well defined samples.

The acquisition of this unprecedented multifrequency multi-satellite data set was used to build well sampled simultaneous SEDs. The SEDs of these sources clearly show the typical two-bump signature usually attributed to Synchrotron and inverse Compton emission. 

The comparison between our simultaneous data with literature archival measurements shows that SEDs built with non-simultaneous data suffer from uncertainties in the $\mu$-wave region that are relatively modest and generally limited to about a factor 2 while the high energy part of the spectrum is much more affected with uncertainties due to flux variations of up to a factor of ten or more. 

\bigskip 
\begin{acknowledgments}
The $Fermi$ LAT Collaboration acknowledges support from a number of agencies and institutes for both development and the operation of the LAT as well as scientific data analysis. These include NASA and DOE in the United States, CEA/Irfu and IN2P3/CNRS in France, ASI and INFN in Italy, MEXT, KEK, and JAXA in Japan, and the K.~A.~Wallenberg Foundation, the Swedish Research Council and the National Space Board in Sweden. Additional support from INAF in Italy and CNES in France for science analysis during the operations phase is also gratefully acknowledged.
\end{acknowledgments}

\end{document}